\documentclass[aip,apl,reprint,amsmath,amssymb]{revtex4-1}
\usepackage{graphicx}
\usepackage{dcolumn}

\begin{document}

\title{Multimode nematicon waveguides}

\author{Yana V. Izdebskaya}
\affiliation{Nonlinear Physics Center, Research School of Physics and Engineering,\\The Australian National University, Canberra, ACT 0200, Australia}

\author{Anton S. Desyatnikov}
\affiliation{Nonlinear Physics Center, Research School of Physics and Engineering,\\The Australian National University, Canberra, ACT 0200, Australia}

\author{Gaetano Assanto}
\affiliation{NooEL--Nonlinear Optics and OptoElectronics Lab, Department of Electronic Engineering, University of Rome ``Roma Tre'', 00146 Rome, Italy}

\author{Yuri S. Kivshar}
\affiliation{Nonlinear Physics Center, Research School of Physics and Engineering,\\The Australian National University, Canberra, ACT 0200, Australia}

\date{\today}

\begin{abstract}
We report on the first experimental observation of higher-order modes guided by soliton-induced waveguides in unbiased nematic liquid crystals. We find that the nematicon waveguides operate in a bounded power region specific to each guided mode. Below this region the guided beams diffract, above it the mode mixing and coupling give rise to an unstable output.
\end{abstract}


\maketitle

Spatial optical solitons propagate in nonlinear Kerr-like media as modes of the self-induced waveguides~\cite{Kivshar}. An attractive property of soliton-induced waveguides is their ability to guide weak signals of different wavelengths~\cite{Shih,Peccianti2000,Petter2001}, as well as the possibility to reconfigure such waveguides by spatial steering~\cite{Petter2001, Peccianti2001, Peccianti2002, Henninot2004}. To this extent, various schemes for all-optical processing have been reported, employing interaction of several solitons~\cite{Peccianti2002,Peccianti2004,Conti2006,Izdebskaya_jeos}, deflection by defects~\cite{Pasquazi2005,Serak2006,Piccardi_2010,Izdebskaya_ol} or patterned nonlinearity~\cite{Beeckman2006,NP06,kaczmarek_inplane}.

Another dimension in soliton-based optical switching is offered by the multi-modal character of self induced waveguides, observed earlier with photorefractive solitons~\cite{Shih,Petter2002} and optical vortices~\cite{PiO}. In this regard, the {\em nematicon} waveguides, i.e. spatial optical solitons in nematic liquid crystals~\cite{Peccianti2000,Peccianti2001,Peccianti2002,Peccianti2004,Izdebskaya_jeos, Pasquazi2005, Serak2006, Conti2006, Piccardi_2010,Izdebskaya_ol,Beeckman2006,NP06}, are of particular interest, because of the long-range or highly nonlocal character of the reorientational nonlinearity. The transverse size of the nematicon-induced index perturbation can be up to one order of magnitude larger than the beam itself~\cite{Conti_2004,Hutsebaut,Henninot2007} and, therefore, such a waveguide is expected to be multi-moded. Additional evidence of such multimodality is the existence of higher-order nonlocal solitons~\cite{1981,PRL07,Beeckman2010} and incoherent solitons~\cite{Buljan2004,Shen2006}, as observed in Refs.~\cite{Opt.Commun,moti05} and~\cite{Mitchell1997,Peccianti_inco}, respectively. Nevertheless, to date, the higher-order modes guided by fundamental nematicons were never observed.

\begin{figure}
\centerline{\includegraphics[width=0.9\columnwidth]{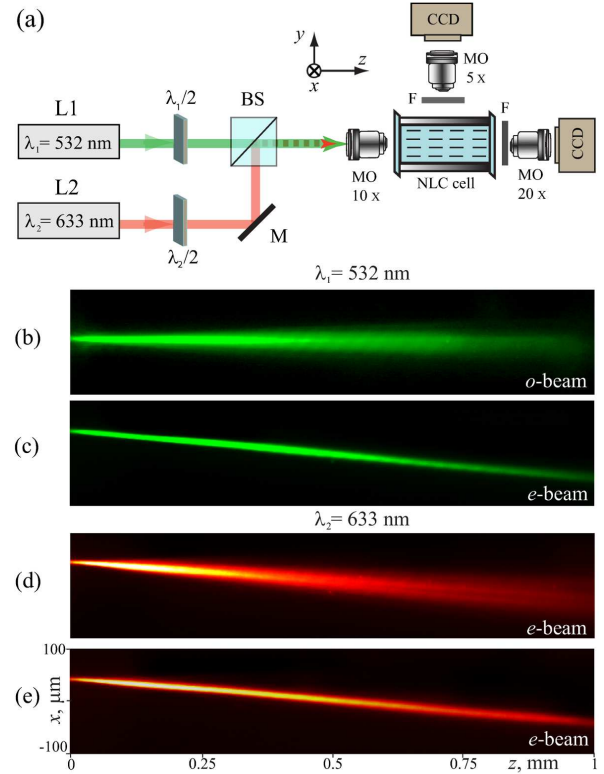}}
\caption{\label{fig1} (Color online) (a) Experimental setup: L1,2 are the green ($\lambda_1=532$~nm) and red ($\lambda_2=633$~nm) cw lasers, $\lambda/2$ - half wavelength plates, BS - beam splitter, M - mirror, MO - micro-objectives, F - filter, CCD - camera. (b-e) Top-view images of (b) ordinary and (c,d,e) extraordinary beams for (b,c) green beam ($P=2$~mW), (d) diffracting red beam and (e) red beam guided by a nematicon.}
\end{figure}

In this Letter we study experimentally the multimode soliton waveguides in nematic liquid crystals, identifying the power domains where high-order guided modes are supported. These domains are limited from above, because the number of modes grows with nematicon power, whereas their mixing eventually gives rise to unstable outputs. The cell geometry and finite size sets additional limits to the order and orientation of nematicon-guided modes. Because of liquid crystal birefringence, the distinction between hybrid EH and HE modes is not possible~\cite{Snyder}, and we use the notation H$_{mn}$ or Hermite-like modes with indices $(m,n)$.

Experimental setup is shown in Fig.~\ref{fig1}(a). We employ a planar cell with the (propagation) length of 1.1~mm, formed by two $L=100\,\mu$m-spaced parallel polycarbonate slides containing the 6CHBT nematic liquid crystal (NLC). The mean angular orientation of the NLC molecules (the molecular \textsl{director}) was pre-set by mechanically rubbing the internal interfaces at $\pi/4$-angle with respect to $z$ in the $(x,z)$ plane, parallel to the slides. To prevent the formation of exposed menisci, two additional $150\,\mu$m-thick slides were attached perpendicularly to $z$ to seal the cell input and output. We use micro-objectives and CCD cameras to collect the light at the sample output and scattered above the cell during propagation.

Figures~\ref{fig1}(b,c) show sample top-view photos of either ordinary (`{\em o}', electric field parallel to $y$) or extraordinary (`{\em e}', electric field parallel to $x$) polarized green beams ($\lambda=532$~nm) launched with $2$~mW input power and the wavevector $\bf k$ aligned to $z$. Clearly, the {\em o}-beam in Fig.~\ref{fig1}(b) diffracts below the Freedericks threshold, whereas the {\em e}-beam in Fig.~\ref{fig1}(c) is self-trapped while walking-off at an angle $\sim5^\circ$ with respect to $\bf k$. The $e$-signal beam (red, $\lambda=633$~nm) is copolarized and launched collinearly with the green beam, resulting in either diffraction without nematicon [Fig.~\ref{fig1}(d)], or guided-wave confinement by nematicon [Fig.~\ref{fig1}(e)].

\begin{figure}
\centerline{\includegraphics[width=1\columnwidth]{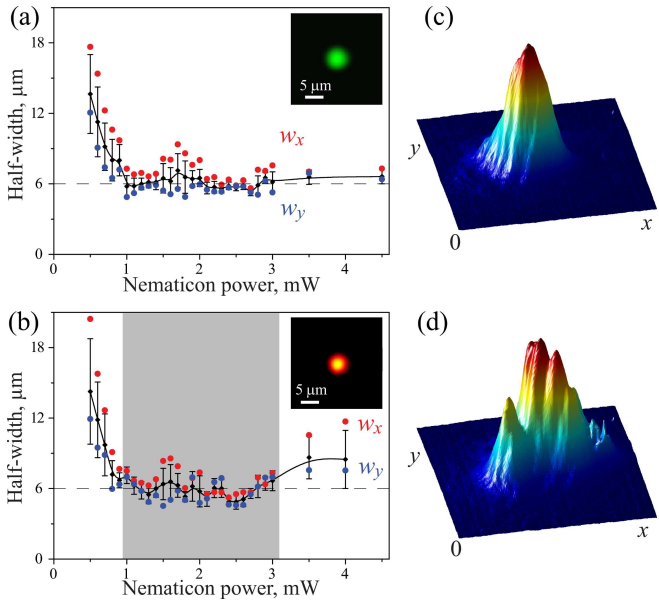}}
\caption{\label{fig2} (Color online) Beam HWHM $w_{x,y}$ (dots) of (a) green nematicon and (b) red  fundamental guided mode; the solid lines with error bars graph the radii of the fitting circular profile. The shaded area in (b) marks the region of stable guiding. The insets in (a) and (b) are photos of the green and red input beams, respectively. (c,d) Averaged intensity distributions of the mode guided by nematicons of powers (c) $P=1.5$~mW and (d) $P=4$~mW.}
%\vspace{-3mm}
\end{figure}

\begin{figure}
\centerline{\includegraphics[width=1\columnwidth]{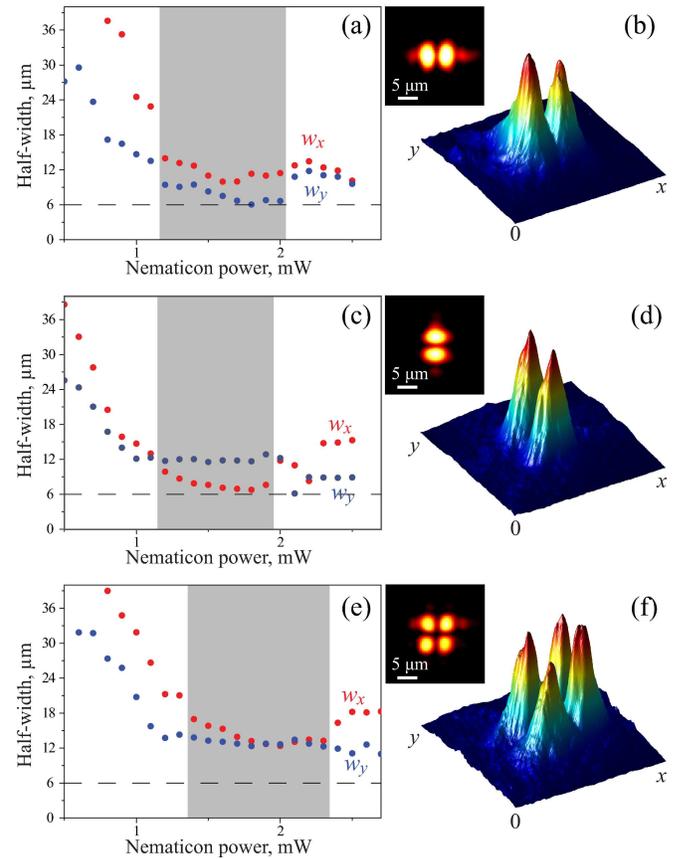}}
\caption{\label{fig3} (Color online) Experimental results at the cell output for (a,b) H$_{10}$, (c,d) H$_{01}$, and (e,f) H$_{11}$ modes launched in the red at the input and guided by a nematicon. (a,c,e) HWHM versus nematicon power $P$; the shaded regions indicate domains of stable guiding. (b,d,f) Averaged output intensity distributions for $P=1.5$~mW. The insets show the input profiles of the red signal.}
%\vspace{-2mm}
\end{figure}

Noteworthy, the nematicons are generated in a uniaxial liquid medium subject to slow dynamics~\cite{time} and instabilities~\cite{Conti2006,MI,Izdebskaya_oe}, where birefringence-induced walk-off is power sensitive~\cite{Izdebskaya_oe,wo3} and nonlocality leads to beam breathing~\cite{Conti_2004,transverse}. In the absence of external bias, as in our configuration~\cite{Izdebskaya_jeos,Izdebskaya_ol,Izdebskaya_oe}, these effects result in a bent non-uniform waveguide with fluctuations on a time scale smaller than the inverse maximum frame-rate of our camera (25 fps). Corresponding changes in modal profiles and the time-dependent mode mixing~\cite{Snyder} demand for temporal averaging of output images recorded during data acquisition, typically 40 frames at 4 fps. Therefore, we define $w_{x,y}$ as the beam half-widths at half-maximum (HWHM) of the averaged intensity distributions in $x$ and $y$ directions, respectively.

Figure~\ref{fig2} shows the experimental results for a green nematicon [Fig.~\ref{fig2}(a)] and a weak red fundamental H$_{00}$ mode co-launched at the input [Fig.~\ref{fig2}(b-d)]. At low power ($P<0.9$~mW) the self-focusing reduces the waist in the green and corresponding size of the guided mode in the red, until a nematicon is generated for $P>0.9$~mW [Fig.~\ref{fig2}(a)] and the red signal becomes a guided mode [Fig.~\ref{fig2}(b)]. The shaded region in Fig.~\ref{fig2}(b) marks the existence of a stable H$_{00}$ guided mode, with typical averaged output transverse profile displayed in Fig.~\ref{fig2}(c). Further increases of nematicon power $P$ lead to a soliton waveguide supporting more guided modes, with a stronger mixing and an output signal exhibiting a multi-hump profile, as in Fig.~\ref{fig2}(d). The latter transition for $P>3$~mW is accompanied by an increase in HWHM for the red signal, as apparent in Fig.~\ref{fig2}(b), whereas the nematicon maintains its robust structure. Clearly, the nonlinear refractive index potential gets higher and wider with nematicon excitation, inducing a multi-moded channel waveguide where a complex pattern appears owing to superposition and mixing of guided modes~\cite{Snyder}.

\begin{figure}
\centerline{\includegraphics[width=1\columnwidth]{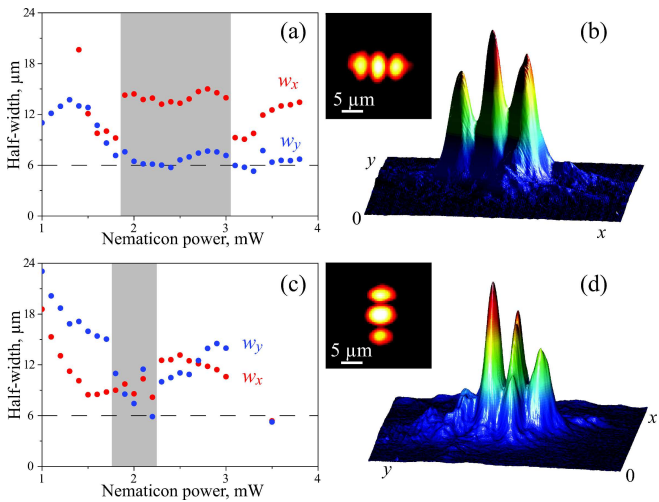}}
\caption{\label{fig4} (Color online) Experimental results at the cell output for (a,b) H$_{20}$ and (c,d) H$_{02}$ modes guided by a nematicon-induced waveguide; (a,c) HWHM vs. nematicon power $P$; (b,d) averaged intensity distributions  for $P=2$~mW. The insets show the input mode profiles in the red.}
\end{figure}

To excite the first-order mode in the soliton-induced waveguide we insert a thin glass plate in front of half of the signal beam and tilt it, introducing a $\pi$-phase jump, to reproduce the phase profiles of H$_{10}$ and H$_{01}$ modes, see the insets in Figs.~\ref{fig3}(a,c). The HWHM of both modes versus nematicon excitation [Figs.~\ref{fig3}(a,c)] exhibit regions of stable guidance as in Figs.~\ref{fig2}(a,b), with averaged profiles shown in Figs.~\ref{fig3}(b,d)]. At variance with soliton waveguides in photorefractive crystals, subject to a directional bias~\cite{Petter2002}, both dipole-like modes coexist in the same interval of excitations, $1.2 < P \text{[mW]} \leq 2$, suggesting that the role of the cell boundaries and the related index anisotropy are minimal in our sample~\cite{Conti_PRE}. Using two glass plates we also generate and launch H$_{11}$ mode of the red signal, as shown in Figs.~\ref{fig3}(e-f). The HWHM of this mode in both $x$ and $y$ directions correspond well to the size of H$_{10}$ and H$_{01}$ modes above, $\sim12\,\mu$m, approximately twice wider than the nematicon (see dashed lines). As expected, the (shaded) stable domain for H$_{11}$ is shifted towards higher powers with $1.4 <P \text{[mW]}< 2.3$.

Finally, Fig.~\ref{fig4} shows the experimental results for H$_{20}$ and H$_{02}$ modes. The H$_{20}$ mode displays a relatively broad domain of stable existence, $1.8 <P \text{[mW]}< 3$. Its HWHM $w_x\backsimeq 15\,\mu$m is nearly three times larger than the nematicon. In sharp contrast, the vertically oriented H$_{02}$ mode [Fig.~\ref{fig4}(d)] is only stable in a narrow domain, $1.7 <P \text{[mW]}< 2.2$, due to the (anisotropic) role of the cell boundaries in $y=0,L$ with fixed director orientation, and despite the fact that its total width, $2w_y<24\,\mu$m, is much smaller than the cell thickness $L=100\,\mu$m. It follows that the order of the supported guided modes is limited from above not only by the nematicon power and mode-mixing process, but also by the boundary-induced anisotropy~\cite{Conti_PRE}. For H$_{mn}$  with $(m,n)\leqslant 1$, the nematicon waveguides can be considered isotropic and circularly symmetric.

In conclusion, we have shown experimentally that the waveguides induced by spatial solitons in unbiased nematic liquid crystals can support various co-polarized guided modes. Multimode waveguides exist in well defined regions of excitations. The confinement of several guided modes in self-induced waveguides is potentially useful for soliton-based optical switches, interconnects and all-optical processing.

The authors acknowledge the Australian Research Council for financial support and thank Prof. W. Krolikowski and Dr. A. Alberucci for useful suggestions.


\begin{thebibliography}{99}
		
\bibitem{Kivshar} Yu. S. Kivshar and G. P. Agrawal, {\em Optical Solitons: From Fibers to Photonic Crystals} (Academic Press, San Diego, 2003).

\bibitem{Shih} M.-F. Shih,  M. Segev, and G. Salamo, Opt. Lett. {\bf 21}, 931 (1996).

\bibitem{Peccianti2000} M. Peccianti, G. Assanto, A. De Luca, C. Umeton, and I. C. Khoo, Appl. Phys. Lett. {\bf 77}, 7 (2000).

\bibitem{Petter2001} J. Petter and C. Denz, Opt. Commun. {\bf 188}, 55 (2001).

%%% steering nematicon + signal

\bibitem{Peccianti2001} M. Peccianti and G. Assanto, Opt. Lett. {\bf 26}, 1690 (2001).

%%% interaction

\bibitem{Peccianti2002} M. Peccianti, K. A. Brzdkiewicz, and G. Assanto, Opt. Lett. {\bf 27}, 1460 (2002).

\bibitem{Henninot2004} J. F. Henninot, M. Debailleul, R. Asquini, A. d'Alessandro, and M. Warenghem, J. Opt. A {\bf 6}, 315 (2004).

\bibitem{Peccianti2004} M. Peccianti, C. Conti, G. Assanto, A. De Luca, and C. Umeton, Nature {\bf 432}, 733 (2004).

\bibitem{Conti2006} C. Conti, M. Peccianti, and G. Assanto, Opt. Lett. {\bf 31}, 2030 (2006).

\bibitem{Izdebskaya_jeos} Ya. V. Izdebskaya, V. Shvedov, A. S. Desyatnikov, W. Krolikowski, G. Assanto, and Yu. S. Kivshar, J. Eur. Opt. Soc. Rapid Publ. {\bf 5}, 10008 (2010).

%%% defect

\bibitem{Pasquazi2005} A. Pasquazi, A. Alberucci, M. Peccianti, and G. Assanto, Appl. Phys. Lett. {\bf 87}, 261104 (2005).

\bibitem{Serak2006} S. V. Serak, N. V. Tabiryan, M. Peccianti, and G. Assanto, IEEE Photon. Tech. Lett. {\bf 18}, 1287 (2006).

\bibitem{Piccardi_2010}  A. Alberucci, A. Piccardi, U. Bortolozzo, S. Residori, and G. Assanto, Opt. Lett {\bf 35}, 390 (2010).

\bibitem{Izdebskaya_ol} Ya. V. Izdebskaya, V. G. Shvedov, A. S. Desyatnikov, W. Z. Krolikowski, and Yu. S. Kivshar, Opt. Lett. {\bf 35}, 10 (2010).

%%% pattern

\bibitem{Beeckman2006} J. Beeckman, K. Neyts, and M. Haelterman, J. Opt. A {\bf 8}, 214 (2006).


\bibitem{NP06} M. Peccianti, A. Dyadyusha, M. Kaczmarek, and G. Assanto, Nat. Physics {\bf 2}, 737 (2006).

\bibitem{kaczmarek_inplane} A. Piccardi, M. Peccianti, G. Assanto, A. Dyadyusha,  and M. Kaczmarek, Appl. Phys. Lett. {\bf 94}, 091106 (2009).
%%%%%


\bibitem{Petter2002} J. Petter, C. Denz, A. Stepken, and F. Kaiser, J. Opt. Soc. Am. B {\bf 19}, 1145 (2002).

\bibitem{PiO} A. S. Desyatnikov, Yu. S. Kivshar, and L. Torner, Prog. Opt. {\bf 47}, 291 (ed. E. Wolf, North-Holland, Amsterdam, 2005).

%\bibitem{Khoo} I. C. Khoo, {\em Liquid crystals: Physical Properties and Nonlinear Optical Phenomena} (Wiley, New York, 1995).

%%%% induced waveguide

\bibitem{Conti_2004} C. Conti, M. Peccianti and G. Assanto, Phys. Rev. Lett. {\bf 92}, 113902 (2004).

\bibitem{Hutsebaut} X. Hutsebaut, C. Cambournac, M. Haelterman, J. Beeckman, and K. Neyts, J. Opt. Soc. Am. B {\bf 22}, 1424 (2005). %phase-measurement interferometry

%\bibitem{Warenghem2006} M. Warenghem, J. F. Blach, and J. F. Henninot, Mol. Cryst. Liq. Cryst. {\bf 454}, 297/[699] (2006).

%Raman scattering

\bibitem{Henninot2007} J. F. Henninot, J. F. Blach, and M. Warenghem, J. Opt. A {\bf 9}, 20 (2007). % interaction and guiding

%%% higher-order solitons

\bibitem{1981} V. A. Mironov, A. M. Sergeev, and E. M. Sher, Sov. Phys. Dokl. {\bf 26}, 861 (1981).

\bibitem{PRL07} D. Buccoliero, A. S. Desyatnikov, W. Krolikowski, and Yu. S. Kivshar, Phys. Rev. Lett. {\bf 98}, 053901 (2007).

%\bibitem{SPRL07} S. Skupin, M. Saffman, and W. Krolikowski, Phys. Rev. Lett. {\bf 98}, 263902 (2007).

\bibitem{Beeckman2010} J. Beeckman, K. Neyts, P. J. M. Vanbrabant, R. James, and F. A. Fernandez, Opt. Express {\bf 18}, 3311 (2010).

\bibitem{Buljan2004} H. Buljan, T. Schwartz, M. Segev, M. Soljacic, and D. N. Christodoulides, J. Opt. Soc. Am. B  {\bf 21}, 397 (2004).
%\bibitem{ChristoJOSAB} K. G. Makris, H. Sarkissian, D. N. Christodoulides, and G. Assanto, J. Opt. Soc. Am. B {\bf 22}, 1371 (2005).
\bibitem{Shen2006} M. Shen, J. Shi, and Q. Wang, Phys. Rev. E {\bf 74}, 027601 (2006).
%\bibitem{Lopez2007} S. Lopez-Aguayo, and J. C. Gutiérrez-Vega, Phys. Rev. A {\bf 76}, 023814 (2007).

\bibitem{Opt.Commun} X. Hutsebaut, C. Cambournac, M. Haelterman, A. Adamski, and K. Neyts, Opt. Commun. {\bf 233}, 211 (2004).

\bibitem{moti05} C. Rotschild, O. Cohen, O. Manela, M. Segev, and T. Carmon, Phys. Rev. Lett. {\bf 95}, 213904 (2005).

%\bibitem{moti06} C. Rotschild, M. Segev, Z. Xu, Y. V. Kartashov, L. Torner, and O. Cohen, Opt. Lett. {\bf 31}, 3312 (2006).
\bibitem{Mitchell1997} M. Mitchell and M. Segev, Nature {\bf 387}, 880 (1997).

\bibitem{Peccianti_inco} M. Peccianti and G. Assanto, Opt. Lett. {\bf 26}, 1791 (2001).


\bibitem{Snyder} A. M. Snyder and J. D. Love, {\em Optical Waveguide Theory} (Chapmen and Hall, London, 1983).

%%%% power walk-off


%\bibitem{bend1} J. Beeckman, K. Neyts, X. Hutsebaut, C. Cambournac, and M. Haelterman, Opt. Quant. Electron. {\bf 37}, 95-106 (2005).

%\bibitem{bend2} J. Beeckman, K. Neyts, X. Hutsebaut, and M. Haelterman, Opto-Electron. Review {\bf 14}, 263 (2006).
%%%% time
\bibitem{time} J. Beeckman, K. Neyts, X. Hutsebaut, C. Cambournac, and M. Haelterman, IEEE J. Quantum Electron. {\bf 41}, 735 (2005).

\bibitem{MI} M. Peccianti, C. Conti and G. Assanto, Phys. Rev. E {\bf 68}, R025602 (2003).

\bibitem{Izdebskaya_oe} Ya. V. Izdebskaya, V. G. Shvedov, A. S. Desyatnikov, W. Z. Krolikowski, M. Belic, G. Assanto, and Yu. S. Kivshar, Opt. Express {\bf 18}, 3258 (2010).
%\bibitem{wo1} M. Peccianti and G. Assanto, Opt. Lett. {\bf 30}, 2290 (2005).

%\bibitem{wo2} K. Jaworowicza, K. A. Brzdakiewicz, M. A. Karpierz, and M. Sierakowski, Mol. Cryst. Liq. Cryst. {\bf 453}, 301-307 (2006).

\bibitem{wo3} A. Piccardi, A. Alberucci, and G. Assanto, Appl. Phys. Lett. {\bf 96}, 061105 (2010).

%breathing
\bibitem{transverse} M. Peccianti, A. Fratalocchi, and G. Assanto, Opt. Express {\bf 12}, 6524 (2004).

\bibitem{Conti_PRE} C. Conti, M. Peccianti, and G. Assanto, Phys. Rev. E {\bf 72}, 066614 (2005).

%\bibitem{MotiNP} C. Rotschild, T. Schwartz, O. Cohen, and M. Segev, Nat. Photon. {\bf 2}, 371-376 (2008).

\end{thebibliography}
\end{document}